\documentclass[a4paper]{jpconf}
\usepackage{graphicx}
\begin{document}
\title{Three-body Force Effects on the Properties
of Neutron-rich Nuclear Matter}

\author{Wei Zuo}

\address{Institute of Modern Physics, Chinese Academy of Sciences, Lanzhou 730000,
China}

\ead{zuowei@impcas.ac.cn}

\begin{abstract}
We review our research work on the single-particle properties and the equation of state (EOS)
of isospin asymmetric nuclear matter
within the framework of the Brueckner-Hartree-Fock (BHF) approach
extended by including a microscopic three-body force (TBF). The TBF is
shown to affect significantly the nuclear matter EOS and the density dependence of nuclear
symmetry energy at high densities above the normal nuclear matter density, and it is necessary
for reproducing the empirical saturation property of symmetric nuclear matter
in a nonrelativistic microscopic framework.
The TBF-induced rearrangement effect and the ground state (g.s.) correlation effect on the s.p. properties
in neutron-rich nuclear matter are investigated. Both effects turn out to be crucial for predicting
reliably the s.p. properties within the Brueckner framework.
The TBF effect on nucleon superfluidity in neutron star matter and neutron stars has also been
discussed.
\end{abstract}

\section{Introduction}
To determine the equation of state (EOS) and single particle (s.p.) properties
of asymmetric nuclear matter (i.e., neutron-rich nuclear matter)
 in a wide density range is one of the most challenging subjects in
nuclear physics and nuclear astrophysics~\cite{liba:2008,danielewicz:2002,steiner:2005,baldo,fuchs:2006,
chen:2007,toro:2005,glendenning:2000,shapiro:1983}.
The properties of asymmetric nuclear matter and their isospin-asymmetry dependence at relatively low
densities around and below normal nuclear matter density $\rho_0$ play a crucial role in predicting the properties
of neutron-rich nuclei away from the nuclear stability line and heavy nuclei, such as
the radius, the neutron-skin thickness and the density
distribution. In Ref.~\cite{oyamatsu:1998}, it has been shown that the neutron density
distribution of neutron-rich nuclei depends sensitively on the isospin-asymmetry dependence of
the equilibrium density of asymmetric nuclear matter. Theoretical
investigations~\cite{furnstahl:2002,chen:2005a,avancini:2007} have indicated that the neutron
skin thickness is strongly correlated with the slope of symmetry energy around the nuclear
matter saturation density.
Besides the general interest in nuclear physics, the EOS of asymmetric nuclear matter,
especially at supra-saturation densities, is expected to be extremely important for understanding
many observational phenomena in nuclear astrophysics and neutron star
physics~\cite{steiner:2005,glendenning:2000,baldo:2001,lattimer:1991,zuo:2004a,goriely:2007,vidana:2012}. For example,
the EOS of asymmetric nuclear matter is a basic input for neutron star structure
model (TOV equation) and it determines essentially the predicted mass-radius relation and maximum mass of
neutron stars consisting of nucleons and leptons. The density dependence of symmetry energy at high densities
determines the proton fraction in $\beta$-stable (n,p,e,$\mu$) neutron star matter, and thus is
decisive for understanding the cooling mechanism of neutron stars.

Heavy ion collisions (HIC) at intermediate and high energies provide powerful tools in laboratory
for extracting information about the EOS and  s.p. properties of asymmetric nuclear matter.
During the dynamic evolution of HIC
induced by radioactive beams, a transient state of dense and highly
isospin-asymmetric nuclear matter can be formed and the
isospin dynamics and observables have been shown to be quite
sensitive to the isovector part of the EOS and symmetry
potential~\cite{liba:2002,liba:2004,tsang,chen:2005b,liqf:2005,shetty:2007}.
Since the EOS of nuclear matter can not be measured directly in the
experiments of HIC, one has to compare the experimental observables
with the theoretical simulations by using transport models to
extract the information about the nuclear EOS indirectly. The s.p.
potentials felt by protons and neutrons in asymmetric
nuclear medium are basic ingredients of the transport models (such as BUU and QMD models) for HIC.
Therefore, reliable information of the density-,
isospin- and momentum-dependence of the s.p.
potentials in neutron-rich nuclear medium is crucial for
constraining the EOS of asymmetric nuclear matter from the experiments of HIC.
Up to now, the density dependence of symmetry energy at low densities below $\sim 1.2\rho_0$ has been
constrained to a certain extent by the HIC experiments and some structure information of finite nuclei~\cite{liba:2008,tsang,chen:2005b}.
However, the density dependence of symmetry energy at high densities remains poorly known. In 2009, Xiao et al.~\cite{xiao:2009} calculated
the $\pi^-/\pi^+$ ratio in central heavy ion collisions at at SIS/GSI energies by using an isospin- and momentum-dependent
BUU model.  By comparing their calculated results with the experimental data measured by the FOPI Collaboration
at GSI~\cite{reisdorf:2007}, they found that a supper soft density-dependence of symmetry
energy at supra-saturation densities is required for reproducing the FOPI data. In 2010, Feng et al.\cite{feng:2010} did the
similar investigation by adopting an isospin-dependent QMD model and
their results favor a supper stiff symmetry energy at supra-saturation densities
in order to match the FOPI data.
In 2011, Russotto et al.~\cite{russotto:2011} studied the elliptic-flow ratio of neutrons with respect to protons or light complex particles
in $Au + Au$ collisions at 400 MeV/nucleon by using UrQMD model and compared their results with the existing FOPI/LAND data.
They suggested that the density dependence of symmetry energy at high densities is most possibly between soft and stiff.

Theoretically, the EOS and s.p. properties of nuclear matter can be predicted by various many-body approaches
including phenomenological methods and microscopic approaches.
Although almost all theoretical approaches are able to reproduce the empirical value of symmetry energy
at the saturation density,
the discrepancy among the predicted density-dependence of symmetry energy at high densities
by adopting different many-body approaches and/or by using different $NN$ interactions
has been shown to be quite large~\cite{fuchs:2006,chen:2005a,Dieperink:2003,lizh:2006,klahn:2006,gogelein:2009}
For example, within the Skyrme-HF framework, different Skyrme parameters may lead to complete
different and even opposite density-dependence of symmetry energy at supra-saturation density~\cite{chen:2005a}.

Microscopically the EOS and s.p. properties of asymmetric nuclear matter have been
investigated extensively by adopting the Brueckner-Hartree-Fock
(BHF) and the extended BHF
approaches~\cite{bombaci:1991,zuo:1999,zuo:2002b,zuo:2005,zuo:2006}, the relativistic
Dirac-BHF (DBHF) theory~\cite{klahn:2006,dalen:2005,ma:2004,sammarruca:2006},
the in-medium $T$-matrix and Green function
methods~\cite{frick:2005,gad:2007,soma:2008,rios:2009},
and the
variational approach~\cite{akmal:1998,bordbar:2008}.
In the
present paper, we shall review systematically our research work on
the properties of neutron-rich nuclear matter within the framework
of the Brueckner-Bethe-Goldstone (BBG) approach extended to include a
microscopic three-body force(TBF).
The paper is organized
as follows. In section 2, we give an introduction of the
adopted theoretical approaches. The numerical results are reported
and discussed in Section 3. Finally, a summary is given Section 4.

\section{Theoretical Approaches}
Our investigation is based on the BBG theory for asymmetric nuclear
matter~\cite{bombaci:1991,zuo:1999}. The extension of the BBG scheme to
include microscopic three-body forces can be found in
Ref.~\cite{zuo:2002b,grange:1989,zuo:2002a}.
Here we simply give a brief review for completeness. 
The starting point of the BHF approach is
the reaction $G$-matrix, which satisfies the following isospin
dependent Bethe-Goldstone (BG) equation,
\begin{equation}
G(\rho, \beta, \omega )= \nonumber\upsilon_{NN}
+\upsilon_{NN} \sum_{k_{1}k_{2}}\frac{ |k_{1}k_{2}\rangle
Q(k_{1},k_{2})\langle k_{1}k_{2}|}{\omega -\epsilon
(k_{1})-\epsilon (k_{2})}G(\rho, \beta, \omega ) \ ,
\end{equation}
where $k_i\equiv(\vec k_i,\sigma_1,\tau_i)$, denotes the s.p. momentum,
the $z$-component of spin and isospin,
respectively. $\upsilon_{NN}$ is the realistic nucleon-nucleon
($NN$) interaction, $\omega$ is the starting energy. The asymmetry
parameter is defined as $\beta=(\rho_n-\rho_p)/\rho$, where $\rho,
\rho_n$, and $\rho_p$ denote the total, neutron and proton number
densities, respectively.
The BHF s.p. energy spectrum is given by $\epsilon(k)=\hbar^2k^2/(2m) +U_{BHF}(k)$.
In solving the BG equation for the $G$-matrix, the continuous
choice~\cite{jeukenne:1976} is adopted for the s.p. potential $U_{BHF}$ since
it provides a much faster convergence of the hole-line expansion than the gap choice~\cite{song:1998}.
 Under the continuous choice, the s.p. potential describes physically at the lowest BHF level
the nuclear mean field felt by a nucleon in nuclear medium.%~\cite{lejeune:1978}.
The BHF s.p. potential is calculated from the real part of the on-shell $G$-matrix,
\begin{equation} U_{\rm BHF}(k) = \sum_{k'} n(k') {\rm Re} \langle k k'|G(\epsilon(k)
+ \epsilon(k')) |k k'\rangle_A \ . \label{eq:ubhf}
\end{equation}

For the realistic $NN$ interaction $\upsilon_{NN}$, we adopt the
Argonne $V_{18}$ ($AV18$) two-body interaction~\cite{wiringa:1995} plus a
microscopic TBF~\cite{zuo:2002a} constructed by using the meson-exchange current
approach~\cite{grange:1989}.
The parameters of the TBF model
have been self-consistently determined to reproduce the $AV18$
two-body force using the one-boson-exchange potential
model and their values can be found in Ref.~\cite{zuo:2002a}.
 In our calculation, the TBF
contribution has been included by reducing the TBF to an
equivalently effective two-body interaction according to the
standard scheme as described in Ref.~\cite{grange:1989}. In
$r$-space, the equivalent two-body force $V_3^{\rm eff}$ reads:
\begin{eqnarray}\label{eq:tbf}
 \langle \vec r_1^{\ \prime} \vec r_2^{\ \prime}| V_3^{\rm eff} |
\vec r_1 \vec r_2 \rangle = \displaystyle
 \frac{1}{4} {\rm Tr} \sum_{n} \int {\rm d}
{\vec r_3} {\rm d} {\vec r_3^{\ \prime}}\phi^*_n(\vec r_3^{\
\prime}) (1-\eta(r_{13}')) (1-\eta(r_{23}')) \nonumber \\
\times W_3(\vec r_1^{\ \prime}\vec r_2^{\ \prime} \vec r_3^{\
\prime}|\vec r_1 \vec r_2 \vec r_3)
 \phi_n(\vec r_3)
(1-\eta(r_{13}))
 (1-\eta(r_{23})).
\end{eqnarray}
It is worth stressing that the effective force $V_3^{\rm eff}$
depends strongly on density. It is the density dependence of the
$V_3^{\rm eff}$ that induces the TBF rearrangement contribution to
the s.p. properties in nuclear medium within the BHF framework.

At the lowest mean field level, there are two problems for the BHF approach in
predicting the nuclear s.p. properties. First, the lowest-order BHF approximation
destroys the Hugenholtz-Van Hove (HVH) theorem and, at densities around the saturation
density, the predicted optical potential depth turns out to be too deep as compared to
the empirical value~\cite{jeukenne:1976}. This problem can be solved by taking into
 account the effect of ground state (g.s.) correlations~\cite{jeukenne:1976,baldo:1988}.
Second, in Ref.~\cite{danielewicz:2000} it has been pointed out that, at high densities
and high momenta, the BHF potential is too attractive and its momentum dependence
turns out to be too weak for describing the experimental elliptic flow data.
In order to predict reliably and realistically the s.p. properties within
 the Brueckner framework, we have improved the calculation of the s.p. potential
 in asymmetric nuclear matter by going beyond the lowest BHF approximation in two aspects.
 First, we have extended the calculation of the effect of g.s. correlations to
 the case of asymmetric nuclear matter~\cite{zuo:1999}. Second, we have included
 the TBF-induced rearrangement contribution in calculating
 the s.p. properties~\cite{zuo:2005,zuo:2006}. The s.p. potential
 in nuclear medium can be derived from the
functional variation of the potential energy density with respect to
the occupation probability of s.p. states~\cite{ring}, i.e.,
\begin{eqnarray}
U(k) = \frac{\delta E_V }{\delta n_k} =  \sum_{k_1} n_{k_1} \langle
k k_1| G | k k_1\rangle_A + \frac{1}{2} \sum_{k_{1} k_{2}} n_{k_1}
n_{k_2} \langle k_1 k_2 | \frac{\delta G} {\delta n_k} | k_1
k_2\rangle_A ,
\label{eq:uren}
\end{eqnarray}
where the first term in the right hand side corresponds to the
standard BHF s.p. potential. In the case without including the TBF,
the above equation becomes identical with the hole-line expansion of
the mass operator~\cite{jeukenne:1976}. By the aid of the BG
equation, the second term can be
worked out explicitly:
\begin{eqnarray} \frac{\delta G} {\delta
n_k} =
 \frac{\delta V_3^{\rm eff}} {\delta n_k} + G
\frac{Q}{e_{12}} \frac{\delta V_3^{\rm eff}} {\delta n_k} +
\frac{\delta V_3^{\rm eff}} {\delta n_k} \frac{Q}{e}G + G
\frac{Q}{e_{12}}\frac{\delta V_3^{\rm eff}} {\delta n_k}\frac{Q}{e}G
+ G \frac{\delta (Q/e_{12})} {\delta n_k} G .\label{eq:tbfren1}
\end{eqnarray}
In the right hand side, the last term
comes from the density dependence of the effective interaction
$G$-matrix and it leads to the g.s. correlation effect on
the s.p. potential~\cite{jeukenne:1976}. The lowest
contribution of the last term corresponds to the core polarization
(also called Pauli rearrangement) which affects mainly the s.p.
potential around and below the Fermi surface.  The first four
terms arise from the effective force $V^{\rm eff}_3$ which is an
equivalent effective two-body interaction of the TBF and depends
strongly on density. In the four terms of the TBF rearrangement
contribution, the first one is predominated over the other three
ones which contain the interaction matrix elements between two particle
states (unoccupied) and two hole
states (occupied) and are negligible as compared to the first term.
Accordingly, the TBF-induced rearrangement contribution to the s.p.
potential can be calculated as follows~\cite{zuo:2006},
\begin{equation}
U_{\rm TBF}(k) \approx \frac{1}{2} \sum_{k_1 k_2} n_{k_1} n_{k_2}
\left\langle k_1 k_2 \left| \frac{\delta V_3^{\rm eff}} {\delta n_k}
\right | k_1 k_2 \right\rangle_A \label{eq:utbf}
\end{equation}

\section{Results and Discussions}
\subsection{EOS of symmetric nuclear matter and TBF effect}

As well known, the nonrelativistic model of rigid nucleons interacting via realistic two-body forces
fitting in-vacuum nucleon-nucleon scattering data misses
empirical saturation properties of nuclear matter~\cite{coester:1970,lizh:2006}. In order to describe reasonably the nuclear
saturation properties within the nonrelativistic BHF framework, one has to take into account the TBF effect.
\begin{figure}
%\begin{center}
\hspace{6pc}\begin{minipage}{26pc}
\includegraphics[width=26pc]{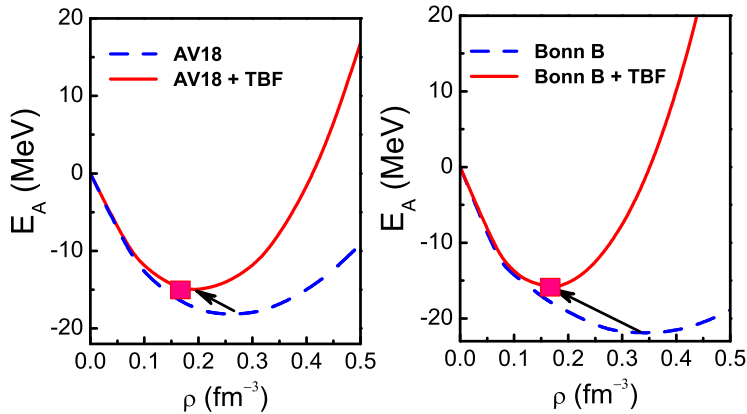}
\end{minipage}
%\end{center}
\caption{\label{fig1} EOS of symmetric nuclear matter. Left panel: the dashed curve is obtained by using the $AV18$ two-body
interaction alone; the solid one by the $AV18$ plus its corresponding TBF. Taken from Ref.~\cite{zuo:2002a}.
Right panel: the dashed curve is obtained by using the $Bonn B$ two-body
interaction alone; the solid one by the $Bonn B$ plus its corresponding TBF. Taken from Ref.~\cite{lizh:2008}.}
\end{figure}

In the left panel of Fig.~\ref{fig1} is shown the
energy per nucleon of symmetric nuclear matter vs. density~\cite{zuo:2002a}, where the dashed curve is obtained by using the $AV18$ two-body
interaction alone; the solid curve is the result by adopting the $AV18$ plus the corresponding TBF.
By comparing the dashed curve and the solid curve, it is seen that the TBF
contribution to the EOS is repulsive and leads to a stiffening of the EOS, especially at supra-saturation densities.
At low densities, the TBF contribution turns out to be
fairly small. The repulsive effect of the TBF increases
monotonically and rapidly as a function density, especially at high densities.  The repulsive contribution
from the TBF turns out improves remarkably the predicted
saturation density of symmetric nuclear matter from $\sim 0.26$fm$^{-3}$ to $\sim 0.19$fm$^{-3}$, indicating that TBF is necessary
for reproducing the empirical saturation property of nuclear matter in a non-relativistic microscopic framework.
In Ref.~\cite{lizh:2008}, we have constructed a new microscopic TBF based on $Bonn B$ two-body interaction. The calculated results by using
the Bonn B interaction and the corresponding TBF are presented in the right panel of Fig.\ref{fig1}. The effect of the TBF based on
the Bonn B potential is seen to be similar to that of the TBF based on $AV18$ potential. In the case of not including the TBF, the
predicted saturation density and the saturation energy are seen to be $\sim 0.33$fm$^{-3}$ and $\sim -22$ MeV, far away from their
empirical values. By adopting the $Bonn B$ interaction plus the corresponding TBF,
the saturation density and the saturation energy are $\sim 0.167$fm$^{-3}$ and $\sim -15.9$ MeV respectively, in satisfactory
agreement with the empirical values. For both TBFs based on the $AV18$ and $Bonn B$ interactions, the connection between
the relativistic effect in the
DBHF approach and the TBF effect has been investigated~\cite{zuo:2002a,lizh:2008}. It has been found that the main relativistic correction
to the EOS of nuclear matter in the DBHF approach can be reproduced
quantitatively by the $2\sigma$-$N\bar{N}$ component of the
microscopic TBF.

\subsection{ EOS of asymmetric nuclear matter and the density dependence of symmetry energy}

\begin{figure}[h]
\begin{minipage}{18pc}\includegraphics[width=18pc]{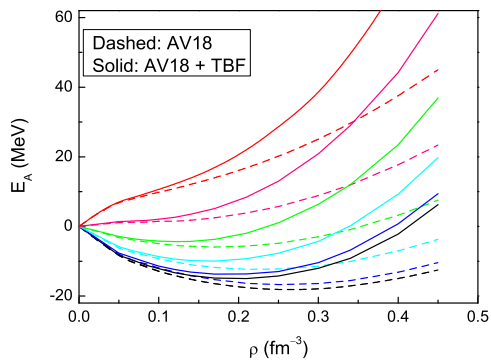}\end{minipage}\hspace{2pc}%
\begin{minipage}[]{14pc}\caption{\label{fig2} EOS of asymmetric nuclear matter at several different isospin-asymmetries,
obtained in the two cases with (solid curves) and without (dashed curves) including the TBF.
Going from the bottom to the top, the asymmetry changes from 0 to 1 in a step of 0.2.}
\end{minipage}
\end{figure}

In Fig.\ref{fig2} is shown the EOS of asymmetric nuclear matter at several different isospin-asymmetries.
The dashed curves denote the results without including the TBF; the solid ones are
obtained by adopting the $AV18$ two-body interaction plus the TBF. From the bottom to the
top in the figure, the corresponding asymmetry parameters are $\beta=0, 0.2, 0.4, 0.6, 0.8$ and 1, respectively
By comparing the dashed curves with the corresponding solid ones, it may be noticed that the TBF gives a
repulsive contribution to the EOS in the whole asymmetry range ($0\ge\beta\ge1$) and inclusion
of the TBF makes the EOS of asymmetric nuclear matter at high asymmetries and/or large densities much stiffer as compared
to the case of not including the TBF. At any given asymmetry, the repulsive TBF contribution increases
 monotonically as a function of density. At a fixed density, the TBF effect turns
out to be more pronounced at a higher asymmetry.

\begin{figure}[h]
\hspace{2pc}\begin{minipage}{18pc}
\includegraphics[width=18pc]{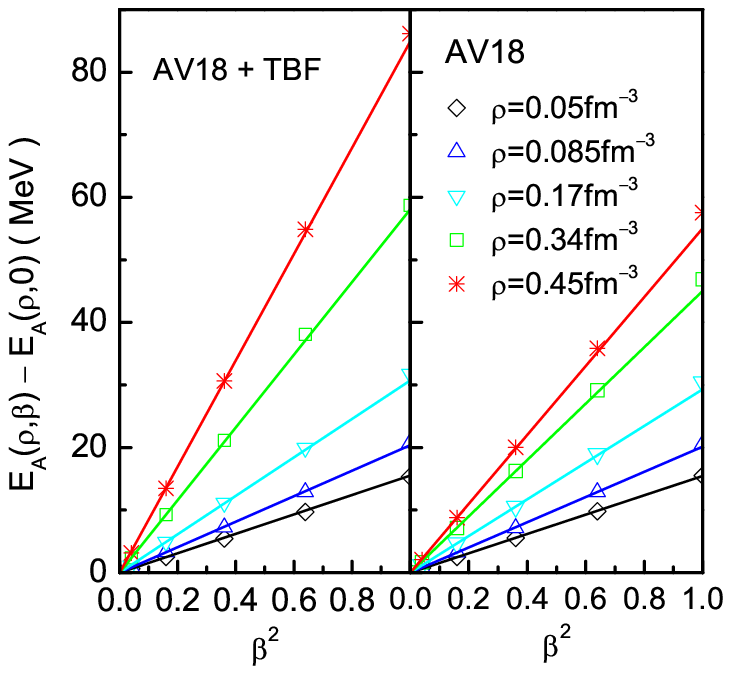}
\caption{\label{fig3} The predicted isovector part of the EOS of asymmetric nuclear matter. Take from Ref.~\cite{zuo:2002b}.}
\end{minipage}\hspace{2pc}%
\begin{minipage}{14pc}
\includegraphics[width=14pc]{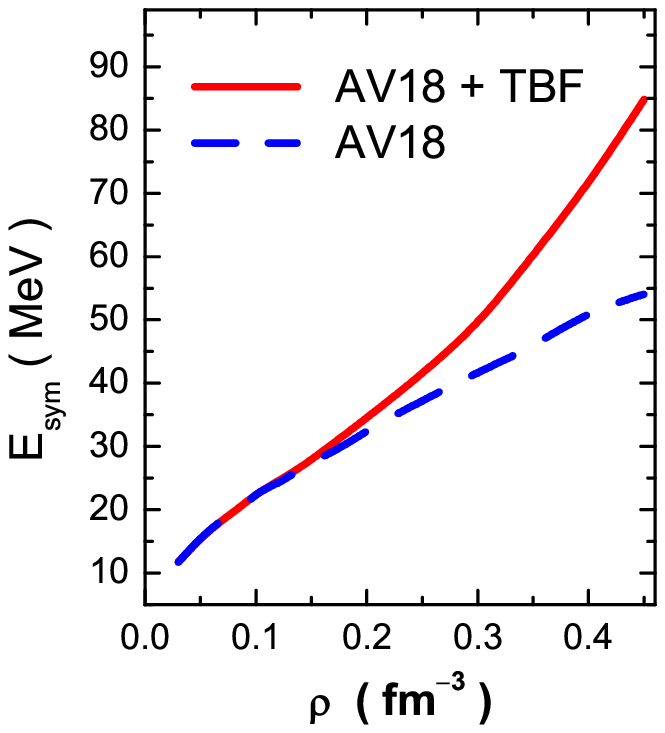}
\caption{\label{fig4} Symmetry energy as a function of density. Taken from Ref.~\cite{zuo:2002b}. }
\end{minipage}
\end{figure}

In order to show more clearly the isospin dependence of the EOS of asymmetric nuclear matter, we report
in Fig.~\ref{fig3} the isovector part of the EOS~\cite{zuo:2002b} (i.e., the difference between the energy per
nucleon of asymmetric nuclear matter at a given $\beta$ and the energy per nucleon
of symmetric nuclear matter) as a function of
$\beta^2$ for several typical densities $\rho=0.085, 0.17, 0.34$ and 0.45fm$^{-3}$.
The right panel corresponds the results without including the TBF; the left panel displays
the results obtained by including the TBF. One may notice from the figure that
the energy per nucleon $E_{A}(\rho,\beta)$ of asymmetric nuclear matter fulfills satisfactorily a linear
dependence on $\beta^2$ in the whole asymmetry
range of $0\leq\beta\le1$. Inclusion of the TBF contribution does not destroy the linear
dependence of $E_{A}(\rho,\beta)$ on $\beta^2$ (see the left panel). Such a linear
dependence of $E_{A}(\rho,\beta)$ on $\beta^2$ is called $\beta^2$-law, which indicates that
the EOS of asymmetric nuclear matter can be expressed as $E_{A}(\rho,\beta)= E_{A}(\rho,0) +
E_{sym}(\rho) \beta^2$ and the higher order terms are negligible. In the above expression,
 $E_{sym}(\rho)$ is the symmetry energy and defined generally as
 $E_{sym}(\rho) = \frac{1}{2}(\partial^2 E_A/\partial \beta^2)_{\beta=0}$. The $\beta^2$-law 
 has also been obtained by Gad {\it et al.}~\cite{gad:2007} within the 
 Green function approach.
 Our above result
provides an microscopic support for the empirical
$\beta^2$-law extracted from the nuclear mass table and extended its
validity up to the highest asymmetry.
The $\beta^2$-law leads to two important consequences.
First, it indicates the symmetry energy can be obtained by
the difference between the EOS of pure neutron matter and that of symmetric nuclear
matter, i.e., $E_{sym}=E_A(\rho,\beta=1)-E_A(\rho,\beta=0)$.
Second, the above $\beta^2$-law implies that the difference of the
neutron and proton chemical potentials in $\beta$-stable neutron
star matter is determined by the symmetry energy
 in an explicit way: $\mu_n-\mu_p=4\beta E_{sym}$ and thus the
 symmetry energy plays a crucial role in predicting the
 composition of neutron stars.
In Ref.~\cite{zuo:2004} we have shown that the above-mentioned $\beta^2$-law is also valid at finite temperatures.

Nuclear symmetry energy describes the isovector part of the EOS of asymmetric nuclear matter.
To see the TBF effect on the density dependence of symmetry energy, in Fig.~\ref{fig4} we compare the
predicted symmetry energy vs. density for the two cases of including the TBF (solid curve) and without
including the TBF (dashed curve)~\cite{zuo:2002b}. In the case of not including the TBF, the density dependence
of symmetry energy is quite soft and follows approximately $E_{sym} \propto
\rho^{0.6}$ in the whole density region of $\rho\le0.5$fm$^{-3}$.
The TBF effect is reasonably small at low densities around and below
the saturation density. At supra-saturation density, the TBF effect on symmetry energy is repulsive and results in a stiffening of the density dependence of symmetry energy.
 Inclusion of the TBF makes the stiffness of the symmetry
energy at high densities become remarkably different from that at
low densities, i.e., the density dependence of symmetry energy changes from a soft one (as compared with a linear dependence) to a stiff one. The thermal effect on the symmetry energy has been studied
in Ref.~\cite{zuo:2004}. It is shown that the thermal effect is small for temperature up to $20$MeV
and as the temperature increases the symmetry
energy decreases slightly.

\subsection{neutron and proton s.p. potentials in asymmetric nuclear matter}

As pointed in Sect.2, within the Brueckner framework extended to include the microscopic
TBF, the full s.p. potential includes three parts:
\begin{equation}
U(k) = U_{\rm BHF}(k) + U_{\rm 2}(k) + U_{\rm TBF}(k).
\label{eq:utot}
\end{equation}
where the first part $U_{\rm BHF}$
corresponds to the lowest-order BHF s.p. potential. The second term $U_2$
describes the effect of the g.s. correlations on s.p. potential and
is called the Pauli rearrangement contribution of $G$-matrix~\cite{jeukenne:1976},
which has been investigated
extensively in literature~\cite{zuo:1999,frick:2005,
gad:2007,soma:2008,rios:2009,jeukenne:1976,baldo:1988}.
The third term $U_{\rm TBF}$ denotes the rearrangement contribution induced by the TBF,
i.e., Eq.(\ref{eq:utbf}).
\begin{figure}[h]
%\begin{center}
\hspace{6pc}\begin{minipage}{26pc}\includegraphics[width=26pc]{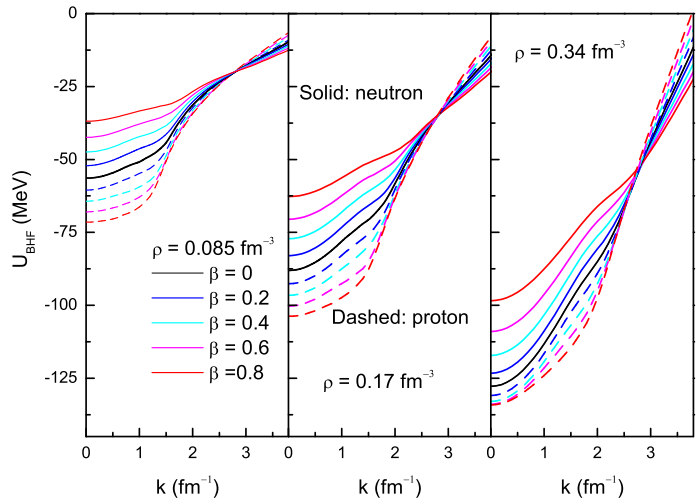}\end{minipage}
%\end{center}
\caption{\label{fig5} Neutron and proton s.p. potentials $U_{\rm BHF}^n$ and $U_{\rm BHF}^p$, predicted at the lowest-order
BHF approximation, in asymmetric nuclear matter at various asymmetries of $\beta=0, 0.2, 0.4, 0.6$ and 0.8 for three typical
densities $\rho=0.085, 0.17$ and 0.34fm$^{-3}$. The solid curves correspond to the neutron s.p. potentials; the dashed
ones denote the proton potentials. Taken from Ref.~\cite{zuo:2005}.}
\end{figure}

In Fig.~\ref{fig5} we display the lowest-order BHF s.p. potentials $U_{\rm BHF}$ for
neutrons and protons in asymmetric nuclear matter at several typical densities
and asymmetries~\cite{zuo:1999,zuo:2005}. It is seen that the BHF s.p. potentials
are strongly attractive at low momenta. At a fixed asymmetry, the attraction of the BHF s.p. potentials turns out
to increase with increasing density. In neutron-rich nuclear matter, the neutron and proton potentials are shown to
become different and split from their common value in symmetric nuclear matter.
For a given density, the neutron s.p. potential $U_{\rm BHF}^n$ becomes less attractive while
the proton potential becomes more attractive at a higher asymmetry. The above predicted different
behavior of neutron and proton
potentials vs. asymmetry $\beta$ is mainly caused by the isospin-singlet $T=0$
$SD$ tensor component of the $NN$ interaction and is readily understood as follows. According to the experimental data on
the phase shifts of $NN$ scattering, the $SD$ channel is strongly attractive at low energies. As the neutron excess increases,
the attraction of the $SD$ interaction between two unlike nucleons becomes stronger for protons and
weaker for neutrons at relatively low momenta. The isospin $T=1$ channel contribution is associated with the variations
of the neutron and proton Fermi surfaces in neutron-rich matter. Its effect on the splitting of the neutron and proton potentials
is opposite and much smaller as compared with the $SD$ channel contribution at low
momenta as discussed in Ref.~\cite{zuo:2005}. The attraction
of the $SD$ channel decreases with energy, so that for high enough momenta, the splitting of the neutron and proton potentials
in neutron-rich matter may vanish and even become opposite due to the competition between  the $T=0$ and $T=1$ channel
effects. In Refs.~\cite{zuo:1999,zuo:2005}, it has been found that the lowest-order BHF neutron and proton potentials
in neutron-rich matter with respect to their common
value in symmetric matter fulfills almost a linear dependence on asymmetry $\beta$, which indicates that the symmetry
potential, defined as $U_{sym}=(U^n-U^p)/2\beta$, is almost independent of $\beta$
 at the lowest-order BHF approximation
and provides a microscopic for the validity of the so-called Lane potential~\cite{lane}.
\begin{figure}
\hspace{6pc}\begin{minipage}{24pc}%\begin{center}
\includegraphics[width=24pc]{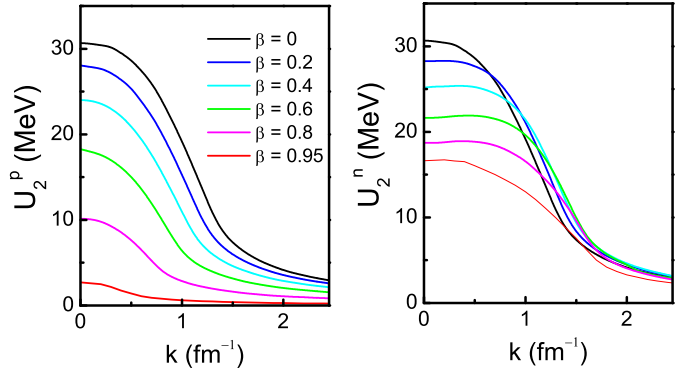}
%\end{center}
\end{minipage}
\caption{\label{fig6} The contributions of g.s. correlations to the neutron and proton s.p. potentials,
$U_2^n$ and $U_2^p$, in neutron-rich nuclear matter at density $\rho=0.17$fm$^{-3}$. Taken from Ref.~\cite{zuo:1999}.}
\end{figure}

In Fig.~\ref{fig6} are plotted the contributions of g.s. correlations to the neutron and proton s.p. potentials
(i.e., $U_2^n$ and $U_2^p$)
at $\rho=0.17$fm$^{-3}$ for several asymmetries $\beta=0, 0.2, 0.4, 0.6$ and 0.8 respectively. In general, it is seen
that the contribution of g.s. correlations is repulsive for both neutron and proton potentials in symmetric nuclear matter
and neutron-rich nuclear matter. The g.s. correlations modifies the neutron and proton potentials mainly at low momenta
around and below the Fermi surfaces and their contribution vanishes rapidly above the corresponding Fermi momenta.
The g.s. correlations are shown to result in a strongly weakening of the momentum dependence of the neutron and proton
s.p. potentials around the respective Fermi surfaces due to the rapid decreasing of their effect as a function of momentum
around the Fermi momenta.
In neutron-rich matter, the $U_2^p(k)$ is found to decrease rapidly as the symmetry $\beta$ increases since the proton Fermi
momentum becomes smaller at a higher asymmetry. For the neutron potential $U_2^n(k)$, the isospin dependence is quite
complicated due to the the coupling between the nucleon hole
states and particle-hole excitations (see the discussion in Ref.~\cite{zuo:1999}). The effect of g.s. correlations may
destroy the linear $\beta$-dependence fulfilled by the neutron and proton potentials at the
lowest-order BHF approximation~\cite{zuo:1999}.
The contribution of the g.s.
correlations not only play an important role in satisfactorily reproducing
the depth of the empirical nuclear optical
potential~\cite{jeukenne:1976}, but is also crucial for restoring the
HVH theorem~\cite{zuo:1999,baldo:1988} and necessary
for generating a nucleon self-energy to describes realistically the
s.p. strength distribution in nuclear matter and finite nuclei below
the Fermi energy~\cite{dickhoff:2004}. However, it can not provide
any appreciate improvement of the high-momentum BHF s.p. potential
which has been shown to be too attractive and
whose momentum-dependence turns out to be too weak at high densities for describing
the experimental elliptic flow data of
HIC experiments at high energies~\cite{danielewicz:2000}.
\begin{figure}[h]
\begin{minipage}{14pc}
\includegraphics[width=16pc]{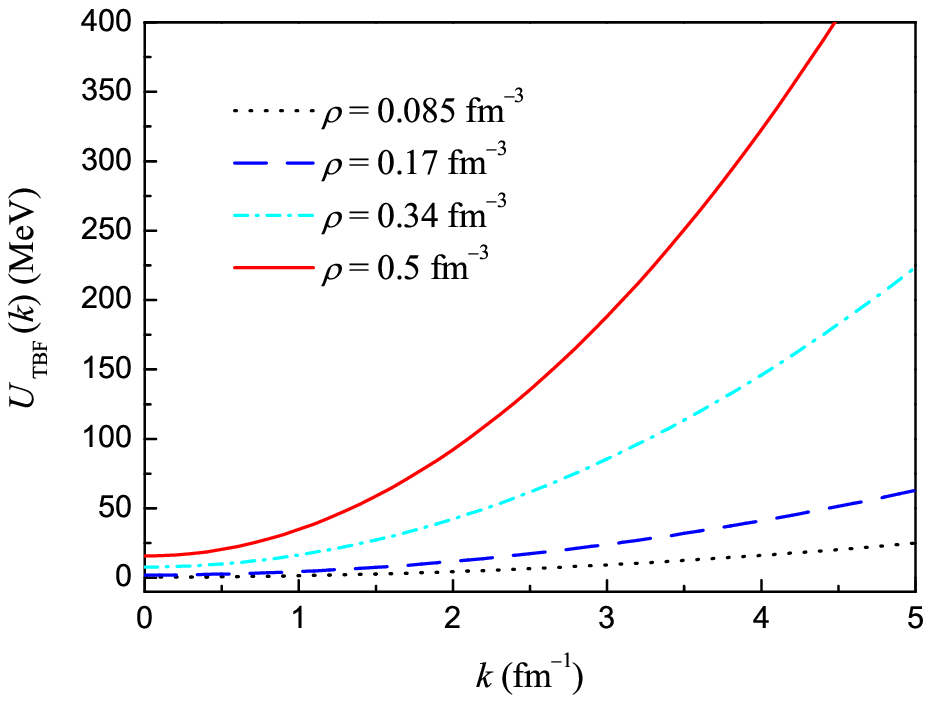}
\caption{\label{fig7} TBF rearrangement contribution $U_{\rm TBF}$ to the s.p. potential in symmetric nuclear matter
for several densities~\cite{zuo:2006}. }
\end{minipage}\hspace{2pc}%
\begin{minipage}{20pc}
\includegraphics[width=20pc]{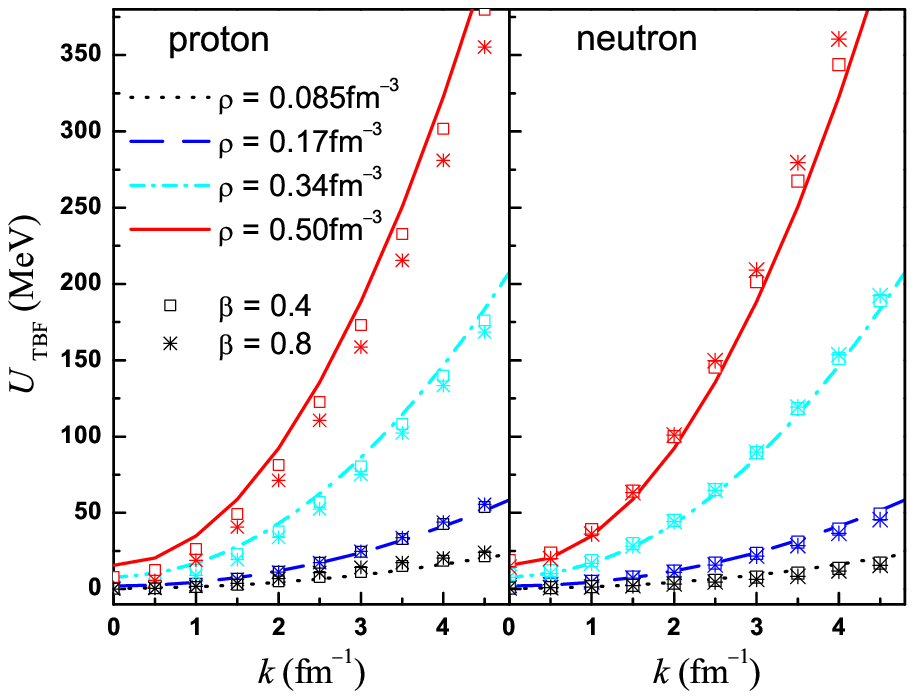}
\caption{\label{fig8} TBF rearrangement contributions $U_{\rm TBF}^p$ and $U_{\rm TBF}^n$ in asymmetric nuclear matter
for $\beta=0.4$ (squares) and $\beta=0.8$ (stars). Taken from Ref.~\cite{zuo:2006}.}
\end{minipage}
\end{figure}

Fig.~\ref{fig7} gives the TBF rearrangement contribution $U_{\rm TBF}$ to
the s.p. potential in symmetric nuclear matter
for several densities~\cite{zuo:2006}.
It is clear from the figure that the TBF rearrangement contribution is
repulsive and its repulsion increases monotonically and rapidly as a function
of density and momentum. One may notice that the rearrangement contribution induced
by the TBF are quite different from
the contribution of g.s. correlations.
At low densities, the TBF rearrangement
contribution $U_{\rm TBF}$ is reasonably small. As the density increases, the $U_{\rm TBF}$ and its
momentum dependence become increasingly stronger.
At high densities and momenta, the TBF induces a
strongly repulsive and momentum-dependent rearrangement
modification of the nucleon s.p. potential in nuclear medium.
It has been shown in Refs.~\cite{zuo:2005,zuo:2006} such a strongly repulsive and momentum-dependent rearrangement
contribution induced by the TBF is crucial for
reducing the disagreement of the large-density and high-momentum
BHF s.p. potential in symmetric matter with
the parametrized potential for describing elliptic flow data~\cite{danielewicz:2000} and those predicted
by the DBHF approach~\cite{dalen:2005}
In neutron-rich nuclear matter, the TBF rearrangement effect is expected to be different on
neutrons from that on protons. To see the isospin dependence, we plotted in Fig.~\ref{fig8}
the TBF rearrangement contribution for neutron and proton ($U_{\rm TBF}^n$ and $U_{\rm TBF}^p$) in
neutron-rich matter~\cite{zuo:2006}. The results for symmetric matter (lines) are also plotted for comparison.
It is seen clearly that in neutron-rich matter both the $U_{\rm TBF}^n$ and $U_{\rm TBF}^p$ are repulsive and
increase rapidly as functions of density and
momentum. The isospin vector parts of the TBF rearrangement
contributions (i.e., the difference between the symbols and
the corresponding lines) turn out to be much smaller in
magnitude than the corresponding isoscalar parts, since the isospin effect in neutron-rich nuclear matter is essentially
a second-order effect in magnitude as compared with the corresponding isoscalar contribution. At sub-saturation densities,
the isospin dependence of the $U_{\rm TBF}^n$ and $U_{\rm TBF}^p$  is seen negligibly small. At supra-saturation densities,
the $U_{\rm TBF}^n$ becomes more repulsive, while the $U_{\rm TBF}^p$ becomes less repulsive at a high asymmetry.

\subsection{neutron and proton effective masses}

Nucleon effective mass describes the momentum dependence of
of the s.p. potential felt by a nucleon in nuclear
medium and is defined as:
\begin{equation}
{m^{\ast\tau}(k) \over m} = \left[{1 + {k \over m}{{\rm d} \, U^{\tau}(k)\over {\rm d} \, k}}\right]^{-1},
\label{eq:mass}
\end{equation}
where $\tau$ denotes neutron or proton. The nucleon effective
mass is of great interest in nuclear physics and nuclear astrophysics~\cite{lunney:2003,goriely:2003,arnould:2007} since it is closely related to
many nuclear phenomena and quantities such as the dynamics of HIC at intermediate and high
energies\cite{liba:2008}, the damping of nuclear excitations and the giant resonances,
the properties of nucleon superfluidity in nuclear matter~\cite{zuo:2008}, $NN$ cross sections
in dense nuclear matter and the transport properties in neutron stars~\cite{zhang:2007},
the nuclear level density around the Fermi surface,
and the physics of stellar collapse~\cite{onsi:2002}.
\begin{figure}[h]
\begin{minipage}{17pc}
\includegraphics[width=17pc]{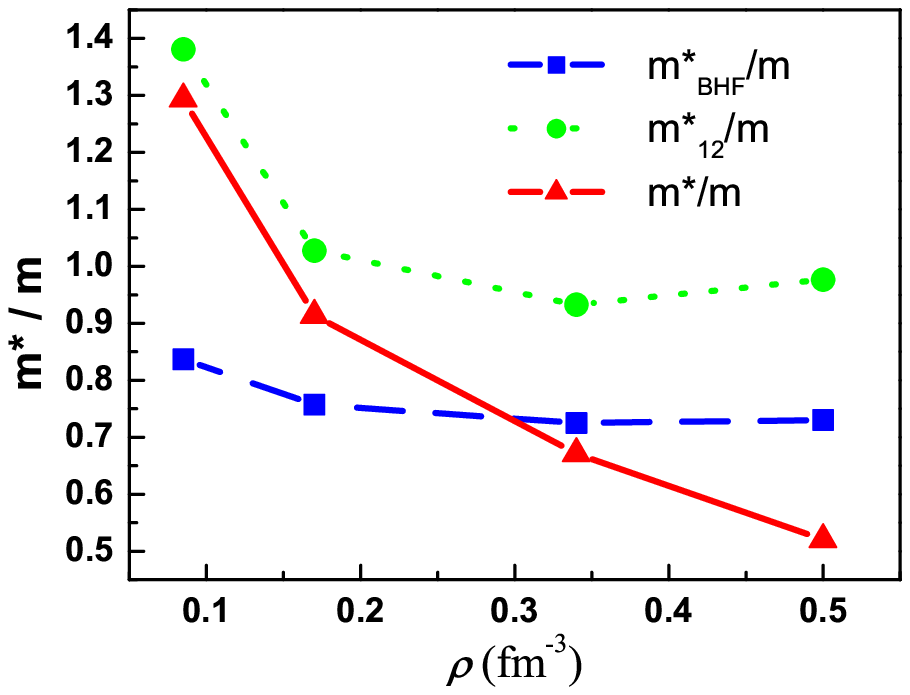}
\caption{\label{fig9}Nucleon effective mass as a function of density at three different level of approximations (see text).
Taken from Ref.~\cite{gan:2011}.}
\end{minipage}\hspace{2pc}%
\begin{minipage}{18pc}
\includegraphics[width=18pc]{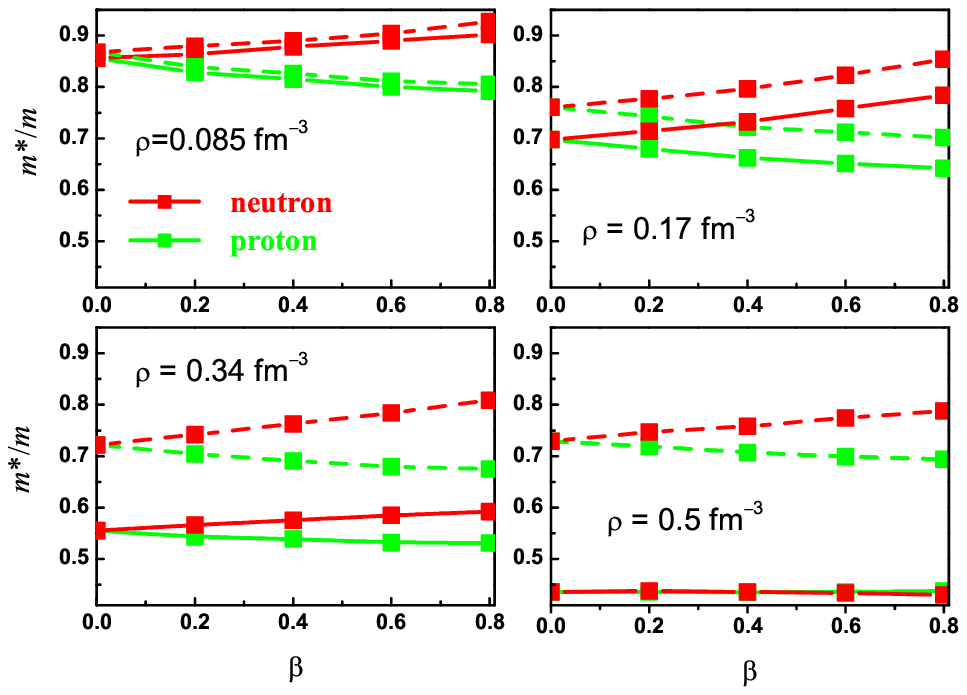}
\caption{\label{fig10}Neutron and proton effective masses vs. $\beta$ in neutron-rich nuclear matter.
The effect of g.s. correlations has not be considered. Taken from Ref.~\cite{zuo:2006}.}
\end{minipage}
\end{figure}

In Fig.~\ref{fig9} we report the calculated effective mass $m^*(k=k_F$) at Fermi momentum vs.
density in symmetric nuclear matter at three different level of approximations:
$m_{\rm BHF}^{\ast}/m$ at the lowest level BHF approximation without including the effect of g.s.
correlations and the TBF rearrangement contribution (dashed curve);
$m^{\ast}_{12}/m$ including the effect of g.s. correlations but without the TBF rearrangement contribution,
i.e., in the Eq.(\ref{eq:mass}) approximating $U$ by
$U \simeq U_{\rm BHF} + U_2$ to get $m^{\ast}_{12}/m$ (dotted curve);
the full effective mass $m^\ast(k)$ by using the full s.p. potential of Eq.~(\ref{eq:utot}).
 At the lowest-order BHF level, the effective mass $m^\ast_{\rm BHF}$ decreases with increasing density
 and saturated at high enough densities. Inclusion of the g.s. correlation effect
leads to significant enhancement of the nucleon effective mass since the g.s. correlations
weakens the momentum dependence of nucleon s.p. potential around Fermi surface~\cite{zuo:1999}.
The TBF rearrangement effect is strongly momentum dependent and consequently it reduces significantly the
nucleon effective mass, especially  at large densities.
One may notice from Fig.~\ref{fig9} that there is a strong competition between the g.s. correlation effect
the TBF rearrangement effect.
Without the TBF rearrangement effect (dashed and dotted curves in the figure), the effective mass decreases
at low densities below and around the saturation density 0.17fm$^{-3}$ and its density dependence
becomes quite weak at high densities.
 By comparing the dashed and dotted curves, it is seen that the g.s. correlations
  result in an overall enhancement of the effective mass in the whole density range up to $\rho=0.5$fm$^{-3}$.
  At low densities, the effective mass is shown to be governed mainly by the g.s. correlation effect
  and inclusion of the TBF rearrangement contribution reduces only slightly the effective mass.
  Inclusion of the TBF rearrangement contribution (solid curve) makes the effective mass become a monotonically
  decreasing function of density in the whole density range considered here.
  At relatively low densities, the full effective mass $m^\ast$ is larger than the lowest order BHF one $m^\ast_{BHF}$,
  while it becomes smaller than the $m^\ast_{BHF}$ at high enough densities,
  which implies that the TBF-induced rearrangement effect becomes predominant over
  the g.s. correlation effect at high enough densities.

In neutron-rich nuclear matter, the neutron and proton effective
masses are expected to split with respect to their common
value in symmetric matter.
In Fig.~\ref{fig10} is depicted the neutron and proton effective masses
at their respective Fermi momenta vs. asymmetry $\beta$ in neutron-rich nuclear matter for several densities $\rho=0.085, 0.17, 0.34$ , and
0.5fm$^{-3}$, respectively. The dashed lines denote the results at the lowest-order BHF approximation; the solid ones are obtained by
including the TBF rearrangement effect. The effect of g.s. correlations is not considered.
It is seen from the figure that both cases with and without the TBF rearrangement contribution,
as the nuclear matter becomes more neutron-rich, the neutron effective mass increases while the
proton one decreases with respect to their
common value in symmetric nuclear matter; i.e., the predicted neutron-proton effective
mass splitting in neutron-rich nuclear matter turns out to be $m^{\ast}_n > m^{\ast}_p$ in
good agreement with
the predictions by the nonrelativistic limit of the DBHF approach~\cite{ma:2004,dalen:2005}.
At the lowest-order BHF approximation, the absolute magnitude of the neutron-proton effective mass
splitting in neutron-rich nuclear matter depends weakly on density and remains almost
the same in the whole density range considered here.
Inclusion of the TBF rearrangement effect makes the
absolute magnitude of the splitting become quite sensitive to the variation of the
density. At sub-saturation densities, the TBF rearrangement effect on the splitting
turns out to be fairly small. At supra-saturation densities,
the effect of the TBF rearrangement leads to a reduction of the neutron-proton effective mass
splitting. At high enough density ($\rho=0.5$fm$^{-3}$), the TBF rearrangement effect may even
suppresses almost completely the splitting.

\subsection{Nucleon superfluidity in neutron star matter and neutron stars}

Nucleon superfluidity plays a crucial role in understanding many astrophysical phenomena
in neutron stars. ~\cite{page:2000,heiselberg:2000,link:2003,gusakov:2005,
pines:1985}. In the inner
crust of a neutron star where the total
 baryon density is low, neutron superfluidity is expected to exist in the
singlet $^1S_0$ channel. In the outer core part, protons may form a superfluid
in the $^1S_0$ partial wave states due to the small proton fraction, and neutron superfluid
may be formed in the $^3PF_2$ coupled channel since the $NN$ scattering data indicate that the
$^3PF_2$ component of the $NN$ interaction is attractive at relatively high energies.
 It is
generally expected that the cooling processes via neutrino emission
\cite{page:2000,heiselberg:2000,link:2003,gusakov:2005}, the magnetic properties,
the properties of rotating dynamics, the post-glitch timing
observations~\cite{shapiro:1983}, and the possible vertex
pinning~\cite{pines:1985} of neutron stars are very sensitive to the
presence of neutron and proton superfluid phases as well as to their
pairing strength.
\begin{figure}[h]
\begin{minipage}{16pc}\includegraphics[width=16pc]{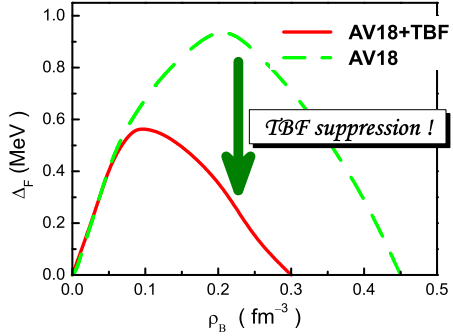}\end{minipage}\hspace{2pc}%
\begin{minipage}[]{14pc}\caption{\label{fig11} Proton
$^1S_0$ pairing gap in $\beta$-stable neutron star matter. The solid curve is
predicted by using the $AV18$ interaction plus the TBF, and the dashed curve by
using the pure $AV18$ two-body force alone. Taken from Ref.~\cite{zuo:2004plb}.}
\end{minipage}
\end{figure}

In Fig.~\ref{fig11} we show the TBF effect on the proton
$^1S_0$ pairing gap $\Delta_F=\Delta(k^n_F)$ in $\beta$-stable neutron star matter predicted within
the BHF + BCS framework~\cite{zuo:2004plb}.
In the figure, the solid curve is predicted by using the AV18 interaction
plus the TBF, and the dashed curve by
using the AV18 two-body interaction alone. As expected, in the case of
not including the TBF, due to the small proton fraction in $\beta$-stable neutron star matter,
the $^1S_0$ proton superfluid
phase may extend to considerably high baryon densities up to $\rho=0.45$fm$^{-3}$
with a peak gap value of 0.95MeV at $\rho\simeq =0.2$fm$^{-3}$.
Inclusion of the TBF leads to a strong suppression of the
$^1S_0$ proton superfluidity in
neutron star matter, especially at high baryon densities.
On the one hand, the TBF reduces significantly the peak
value of the $^1S_0$ proton pairing gap by about 50\%
from $\sim$ 0.95MeV to $\sim$ 0.55MeV and shifts the peak to a much lower baryon
density from $\sim 0.2$fm$^{-3}$ to $\sim 0.09$fm$^{-3}$.
On the other hand, inclusion of the
TBF results in a remarkably shrinking of the density region for the
 $^1S_0$ proton superfluid phase to $\rho<0.3$fm$^{-3}$.
In spite of the small proton fractions in $\beta$-stable neutron star
matter which correspond to small proton densities in the matter,
the above predicted
TBF suppression of the $^1S_0$ proton superfluidity can be readily
understood as follows. Since proton pairs
are embedded inside the medium of neutrons and protons,
both the surrounding protons and neutrons contribute
to the TBF renormalization of the proton-proton pair interaction.
And consequently the relevant density to
the TBF effect on proton pairing is the total baryon density, but not
the proton one, which can be verified from Fig.~\ref{fig11} that
the TBF-induced reduction of the pairing gap becomes stronger at a
higher total baryon density.
The strong weakening
 of the $^1S_0$ proton superfluidity induced by the TBF
 may has important implication for modeling the
neutron-star cooling scenario~\cite{gusakov:2005}.
\begin{figure}[h]
\begin{minipage}{18pc}
\includegraphics[width=18pc]{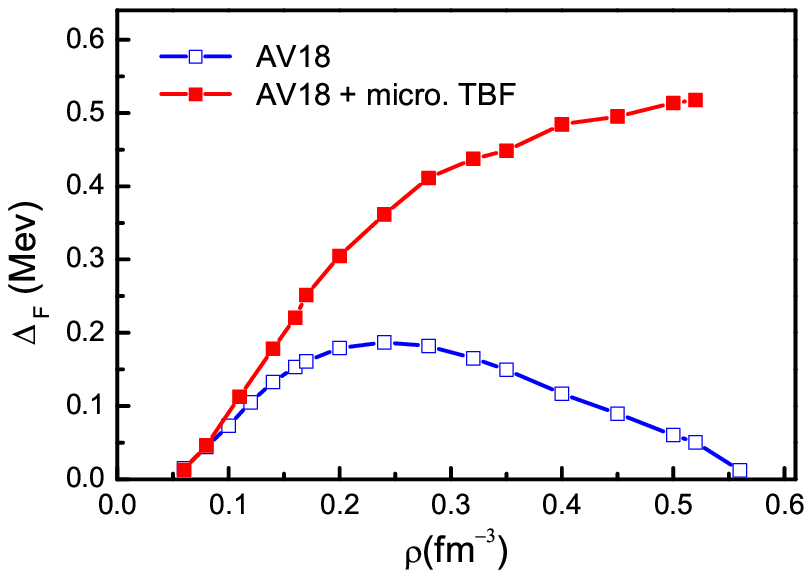}
\end{minipage}\hspace{1pc}%
\begin{minipage}{18pc}
\includegraphics[width=18pc]{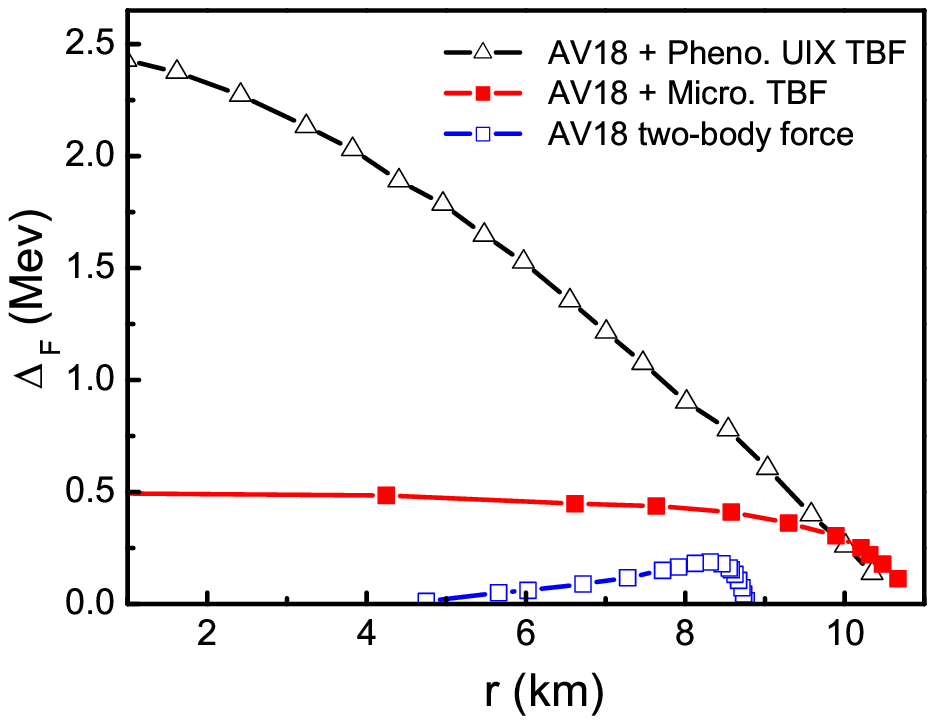}
\end{minipage}
\caption{\label{fig12} Left panel: $^3PF_2$ neutron pairing gap
in $\beta$-stable neutron star matter
 as a function of
density. Eight panel: distribution of the $^3PF_2$ neutron pairing gap
inside a typical neutron star. Taken from Ref.~\cite{zuo:2008}.}
\end{figure}

In Fig.~\ref{fig12} it is shown the $^3PF_2$ neutron pairing gap
in $\beta$-stable neutron star matter
 as a function of the total nucleon
density $\rho$ (left panel) and inside a typical neutron star with a
mass of $M=M_{\odot}$ (right panel)~\cite{zuo:2008}.
In the figure, the empty squares corresponds to the results by using the pure $AV18$ two-body interaction;
the filled squares are predicted by adopting the $AV18$ interaction plus
the microscopic TBF; the empty triangles denote the results obtained by Zhou {\it et al.}~\cite{zhou:2004}
by using the $AV18$ plus the semi-phenomenological UIX TBF.
In the case of not including any TBF,
it is seen from the left panel of Fig.~\ref{fig12} that in $\beta$-stable neutron star matter,
the pairing gap increases first as a function of density, reaching a peak value of
about 0.2~MeV, and then decreases with increasing density in the region of
$\rho\ge0.24$~fm$^{-3}$. By comparing the empty
squares and the filled squares, one may notice that inclusion of the TBF
enhances remarkably the $^3PF_2$ neutron superfluidity in
$\beta$-stable neutron star matter, especially at high densities,
and makes the $^3PF_2$ neutron paring gap become an monotonically
increasing function of density. From the right panel, one may see that the predicted distribution of
the $^3PF_2$ neutron pairing gap in a neutron star by adopting purely the
$AV18$ two-body interaction is similar with the density dependence of the pairing gap in $\beta$ neutron star matter,
i.e., the pairing gap first increases going from the outer part to the inner part of the neutron star,
reaches its maximum value of about 0.2MeV at $R\simeq8.4$km, then starts to decrease and finally vanishes at
$R< 4.7$km in the inner part of the neutron star where the baryon density is expected to be high enough.
By comparing the empty squares and the filled squares, it is clear that inclusion of the
microscopic TBF leads to a significantly overall enhancement of the $^3PF_2$
neutron superfluidity inside neutron stars and makes the superfluid
phase spread throughout the whole neutron star. The microscopic TBF turns out to enlarge the
maximum strength of the pairing gap from $\sim 0.2$ to $\sim 0.5$MeV.
We notice that the semi-phenomenological Urbana UIX TBF leads to an extremely strong $^3PF_2$
neutron superfluidity in neutron stars which is much stronger as compared with our prediction
by adopting the microscopic TBF (for a detailed discussion, see Ref.~\cite{zuo:2008}).

\section{Summary}
We have reviewed our research work on the EOS and the s.p. properties of asymmetric nuclear matter
within the framework of the Brueckner approach extended by including the microscopic TBF.
The TBF is shown to provide a repulsive contribution to the EOS of asymmetric nuclear matter and its repulsion
increases monotonically and rapidly as a function of density and asymmetry at supra-saturation
densities. The repulsive contribution of the TBF leads to a significantly stiffening of the EOS of asymmetric
nuclear matter and the density dependence of symmetry energy at high densities above the normal nuclear matter density,
and it turns out to be necessary for reproducing the empirical saturation property of symmetric nuclear matter
within a nonrelativistic microscopic many-body framework.
The EOS of asymmetric nuclear matter is proved to fulfill satisfactorily a linear dependence on $\beta^2$
in the whole asymmetry range of $0 \le\beta\le 1 $, which supports microscopically the empirical
$\beta^2$-law extracted from the nuclear mass table and extended its
validity up to the highest isospin asymmetry. Inclusion of the
TBF does not destroy the $\beta^2$-law. Inclusion of the TBF makes
the density dependence of symmetry energy at high densities become much stiffer than that at sub-saturation densities.

In predicting the s.p. properties in asymmetric nuclear matter, we have extended and improved the Brueckner approach
in two aspects. One is to extend the calculation of the effect of g.s. correlations to asymmetric nuclear matter;
the second is to include the rearrangement contribution induced by the TBF.
Both the TBF rearrangement contribution and the g.s. correlation effect are shown to be crucial and
necessary for predicting reliably the s.p. properties in neutron-rich nuclear matter within the microscopic Brueckner
framework. The g.s. state correlations give a repulsive contribution to the neutron and proton s.p. potentials,
 and modify mainly the s.p. properties at low momenta around and below the Fermi surfaces. The rapid decreasing
 of the effect of g.s. correlations as a function of momentum around the Fermi momenta $k_F$ may weakening considerably
 the momentum dependence of the s.p. potentials and results in an significant enhancement of the nucleon effective
 masses around $k_F$. Inclusion of the contribution of g.s. correlations destroys
 the linear $\beta$-dependence fulfilled by the neutron and proton potentials at the lowest-order BHF
approximation. The TBF is found to induce a strongly repulsive momentum dependent contribution to the s.p. potentials
at high densities and large momenta. Being different from the effect of g.s. correlations, the TBF-induced rearrangement
repulsion increases rapidly as a function of density and momentum. The TBF rearrangement effect enhances strongly the repulsion
of the s.p. potentials and turns out to be necessary for reducing the disagreement of the large-density and high-momentum
BHF s.p. potential in symmetric matter with the parameterized potential extracted for describing the elliptic flow data in HIC
and those predicted by the DBHF approach. The strong momentum-dependence of the TBF rearrangement
contribution leads to a significant reduction of the nucleon effective mass and makes the effective mass decrease
monotonically and rapidly as a function of density at supra-saturation densities. At sub-saturation densities, the effective mass is
shown to be governed mainly by the g.s. correlation effect. Whereas, the TBF-induced rearrangement effect becomes predominant over
the g.s. correlation effect at high enough densities.
In neutron-rich nuclear matter, the neutron effective mass turns out to be larger than the proton one in both cases with
and without including the TBF rearrangement contribution. The TBF rearrangement effect is shown to reduce remarkably
the isospin splitting of the neutron and proton effective masses in high-density neutron-rich matter.

The TBF effect on nucleon superfluidity in $\beta$-stable neutron star matter and neutron stars has also been
discussed. On the one hand, the TBF is shown to suppress remarkably the $^1S_0$ proton superfluidity in neutron star
matter, especially at high baryon densities.
It reduces significantly the peak
value of the $^1S_0$ proton pairing gap by about 50\%
from $\sim$ 0.95MeV to $\sim$ 0.55MeV and shifts the peak to a much lower baryon
density from $\sim 0.2$fm$^{-3}$ to $\sim 0.09$fm$^{-3}$.
On the other hand, inclusion of the TBF turns out to enhance considerably the predicted $^3PF_2$
neutron superfluidity in neutron star matter and neutron stars.
The microscopic TBF leads to a strongly overall enhancement of $^3PF_2$
neutron superfluidity and makes the corresponding superfluid phase spread throughout
inside the whole inner part of neutron stars.

\section*{Acknowledgments}
The work was supported by the National Natural Science
Foundation of China (11175219, 10875151), the Major State Basic
Research Developing Program of China (No. 2007CB815004), the
Knowledge Innovation Project (KJCX2-EW-N01) of the Chinese Academy of
Sciences, the Chinese Academy of Sciences Visiting Professorship for Senior International
Scientists (Grant No.2009J2-26).


\section*{References}
\begin{thebibliography}{9}

\bibitem{liba:2008} Li B A, Chen L W and Ko C M 2008 {\it Phys. Rep.} {\bf 464} 113 and reference therein

\bibitem{danielewicz:2002} Danielewicz P, Lacey R and Lynch W G 2002 {\it Science} {\bf 298} 1592

\bibitem{steiner:2005} Steiner A W, Prakash M, Lattimer J M et al.
2005 {\it Phys. Rep.} {\bf 411} 325

\bibitem{baldo} Baldo M, Maieron C 2007 {\it J. Phys.} {\bf G34} R243;
 Baldo M, Burgio G F 2012 {\it Rep. Prog. Phys.} {\bf 75} 026301

\bibitem{fuchs:2006} Fuchs C and Wolter H H 2006 {\it Eur. Phys. J} {\bf A 30} 5

\bibitem{chen:2007} Chen L W, Ko C M, Li B A and Yong G C 2007 {\it Front. Phys.
China} {\bf 2} 327

\bibitem{toro:2005} Barana V, Colonna M, Grecoc V and Di Toro M, 2005 {\it Phys. Rep.}
{\bf 410} 335

\bibitem{glendenning:2000} Glendenning N K 2000 {\it Compact Stars:
Nuclear Physics, Particle Physics and General Relativity} (Berlin: Springer)

\bibitem{shapiro:1983} Shapiro S L and Teukolsky S A 1983 {\it Black Holes,
White Dwarfs and Neutron Stars} (New York: Wiley)

\bibitem{oyamatsu:1998} Oyamatsu K, Tanihata I, Sugahara Y et al. 1998
{\it Nucl. Phys.} {\bf A634} 3

\bibitem{furnstahl:2002} Furnstahl R J 2002 {\it Nucl. Phys.} {\bf A706} 85

\bibitem{chen:2005a} Chen L W, Ko C M and Li B A 2005 {\it Phys. Rev} {\bf C72} 064309

\bibitem{avancini:2007} Avancini S S, Marinelli J R, Menezes D P
et al. 2007 {\it Phys. Rev.} {\bf C75} 055805

\bibitem{baldo:2001} Baldo M and Burgio G F 2001 {\it Lect. Notes Phys.} {\bf 578} 1

\bibitem{lattimer:1991} Lattimer J M, Pethick C J, Prakash M
and Haensel P 1991 {\it Phys. Rev. Lett.} {\bf 66}, 2701

\bibitem{zuo:2004a} Zuo W, Li A, Li Z H and Lombardo U 2004
 {\it Phys. Rev.} {\bf C70} 055802

\bibitem{goriely:2007} Goriely S and Delaroche J P 2007 {\it Phys. Lett.} {\bf B653} 178

\bibitem{vidana:2012} Vidana I 2012 {\it Phys. Rev.} {\bf C85} 

\bibitem{liba:2002} Li B A 2002 {\it Phys. Rev. Lett.} {\bf 88} 192701

\bibitem{liba:2004} Li B A, et al. 2004 Phys. Rev. {\bf C69} 011603(R); 2004 {\it Nucl. Phys.} 
{\bf A735} 563; 2006
{\it Phys. Lett.} {\bf B634} 378

\bibitem{tsang} M B Tsang, et al. 2004 {\it  Phys. Rev. Lett.} {\bf 92} 062701;
Tsang M B, et al. 2009 {\it Phys. Rev. Lett.} {\bf 102} 122701


\bibitem{chen:2005b} Chen L W, Ko C M and Li B A 2005 {\it Phys. Rev. Lett.} {\bf 94} 032701

\bibitem{liqf:2005} 
Li Q F, Li Z X, Soff S 2005 {\it J.Phys.} {\bf G31} 1359; 
Li Q F, Li Z X and Stoecker H 2006 {\it Phys . Rev.} {\bf C73} 051601


\bibitem{shetty:2007} Shetty D V, Yennello S J and Souliotis G A 2007 {\it Phys. Rev.} {\bf C76}
024606

\bibitem{xiao:2009} Xiao Z G, Li B A, Chen L W et al. 2009 {\it Phys. Rev. Lett.} {\bf 102} 062502
\bibitem{reisdorf:2007} Reisdorf W, Stockmeier M, Andronic A et al., FOPI Collaboration 2007 {\it Nucl. Phys.}
{\bf A781} 459
\bibitem{feng:2010} Feng Z Q and Jin G M 2010 {\it Phys. Lett.} {\bf B683} 140
\bibitem{russotto:2011} Russotto P, Wu P Z, Zoric M 2011 {\it Phys. Lett.} {\bf B697} 471


\bibitem{Dieperink:2003} Dieperink A E L, Dewulf Y, Van Neck D, Waroquier M and Rodin V 2003
{\it Phys. Rev.} {\bf C68} 064307
\bibitem{lizh:2006} Li Z H, Lombardo U, Schulze H J, Zuo W,
Chen L W and Ma H R 2006
{\it  Phys. Rev.} {\bf C74} 047304
\bibitem{klahn:2006} Klahn T, Blaschke D, Typel S et al. 2006 {\it Phys. Rev.} {\bf C74} 035802
\bibitem{gogelein:2009} G\"{o}gelein P, van Dalen E N E, Gad Kh, Hassaneen Kh S A and M\"{u}ther M 
2009 {\it Phys. Rev.} {\bf C79} 024308

\bibitem{bombaci:1991} Bombaci I, lombardo U 1991 {\it Phys. Rev.}
 {\bf C 44} 1892; Baldo M, et al. 1997 {\it Astron. Astrophys.}
{\bf 328} 274


\bibitem{zuo:1999} Zuo W, Bombaci I and Lombardo U 1999 {\it Phys. Rev.}
{\bf C60} 024605

\bibitem{zuo:2002b}Zuo W, Lejeune A, Lombardo U and Mathiot J F 2002
{\it Eur. Phys. J.} {\bf A14} 469
\bibitem{zuo:2005} Zuo W, Cao L G, Li B A, Lombardo U and
Shen C W 2005 {\it Phys. Rev.} {\bf C72} 014005

\bibitem{zuo:2006} Zuo W, Lombardo U, Schulze H J and
Li Z H 2006 {\it Phys. Rev.} {\bf C74} 014317

\bibitem{dalen:2005} van Dalen E N E, Fuchs C and Faessler A 2005
{\it Phys. Rev. Lett.} {\bf 95} 022302; 2005 {\it Phys. Rev.} {\bf C72} 065803
\bibitem{ma:2004} Ma Z Y, Rong J, Chen B Q et al. 2004 %, Z. Y. Zhu, and H. Q. Song,
{\it Phys. Lett.} {\bf B604} 170
\bibitem{sammarruca:2006} Krastev P and Sammarruca F 2006 {\it Phys. Rev.}
{\bf C73} 014001


\bibitem{frick:2005} Frick T, M\"uther H, Rios A, Polls A and Ramos A 2005
 {\it Phys. Rev.} {\bf C71} 014313
\bibitem{gad:2007} Kh. Gad and Kh. S. A. Hassaneen,
{\it Nucl. Phys.} {\bf A793} (2007) 67.
\bibitem{soma:2008} Soma V and Bozek P 2008 {\it Phys. Rev.} {\bf C78} 054003
\bibitem{rios:2009} Rios A, et al. 2009 {\it Phys. Rev. } {\bf C79} 064308; 
Rios A and Soma V 2012 {\it Phys. Rev. Lett.} {\bf 108} 012501
%Polls A and Dickhoff W H
\bibitem{akmal:1998} Akmal A, Pandharipande V R and Ravenhall D G 1998 {\it Phys. Rev.} {\bf C58} 1804

\bibitem{bordbar:2008} G. H. Bordbar and M. Bigdeli, {\it Phys.
Rev.} {\bf C77} (2008) 015805.


\bibitem{grange:1989} Grang\'e P, Lejeune A, Martzolff M and Mathiot J F 1989
{\it Phys. Rev.} {\bf C40} 1040
\bibitem{zuo:2002a} Zuo W, Lejeune A, Lombardo U and Mathiot J F 2002
{\it Nucl. Phys.} {\bf A706} 418

\bibitem{jeukenne:1976} Jeukenne J P, Lejeune A and
Mahaux C 1976 {\it Phys. Rep.} {\bf 25} 83

\bibitem{song:1998} Song H Q, Baldo M, Giansiracusa G and Lombardo U 1998
{\it Phys. Rev. Lett.} {\bf 81} 1584

\bibitem{wiringa:1995} Wiringa R B, Stoks V G J, and Schiavilla R 1995
{\it Phys. Rev.} {\bf C51} 38

\bibitem{baldo:1988} Baldo M, et al. 1988 {\it Phys. Lett.} {\bf B209} 135;
Baldo M, et al. 1990 {\it Phys. Rev.} {\bf C41} 1748

\bibitem{danielewicz:2000} Danielewicz P 2000 {\it Nucl. Phys.} {\bf A673} 375

\bibitem{ring} Ring P and Schuck P 1980 {\it The Nuclear
  Many Body Problem} (New York: Springer-Verlag)


\bibitem{lizh:2008} Li Z H, Lombardo U, Schulze H J and  Zuo W 2008
{\it Phys. Rev.} {\bf C77} 034316


\bibitem{coester:1970} Coester F, Cohen S, Day B and Vincent C M 1970
{\it Phys. Rev.} {\bf C1} 769

\bibitem{zuo:2004} Zuo W, Li Z H, Li A, et al. 2004 {\it Phys. Rev.} {\bf C69} 064001; 
{\it Nucl. Phys.} {\bf A745} 34

\bibitem{lane} Lane A M 1962 {\bf Nucl. Phys.} {\bf 35} 676


\bibitem{dickhoff:2004} Dickhoff W H and Barbieri C 2004 
{\it Prog. Part. Nucl. Phys.} {\bf 52} 377

\bibitem{zuo:2008} Zuo W, Cui C X, Lombardo U and Schulze H J 2008
{\it Phys. Rev.} {\bf C78} 015805

\bibitem{gan:2011}Gan S X, Zuo W and Lombardo U 2011 {\it Chinese Phys.} {\bf C36} 513 


\bibitem{lunney:2003} Lunney D, Pearson J M and Thibault C 2003
{\it Rev. Mod. Phys.} {\bf 75} 1021

\bibitem{goriely:2003} Goriely S, Bender M, Pearson M et al. 2003
{\it Phys. Rev.} {\bf C68} 054325.

\bibitem{arnould:2007} Arnould M, Goriely S and Takahashi K 2007
{\it Phys. Rep.} {\bf 450} 97


\bibitem{zhang:2007} Zhang H F, Li X H, Lombardo, Luo P Y, Sammarruca F and Zuo W 2007
{\it  Phys. Rev.} {\bf C76} 054001

\bibitem{onsi:2002} Onsi M and Pearson J M 2002 {\it Phys. Rev.}
{\bf C65} 047302


\bibitem{page:2000} Page D, Prakash M, Lattimer J M and Steiner A 2000
 {\it Phys. Rev. Lett.} {\bf 85} 2048

\bibitem{heiselberg:2000} Heiselberg H and Hjorth-Jensen M 2000
 {\it Phys. Rep.} {\bf 328} 237

\bibitem{link:2003} Link B 2003 {\it Phys. Rev. Lett.} {\bf 91} 101101

\bibitem{gusakov:2005} Gusakov M E, et al. 2005
{ \it Mon. Not. R. Astron. Soc.} {\bf 363} 555;
Kaminker A D, et al. 
{\it ibid} {\bf 365} 1300


\bibitem{pines:1985} Pines D and Alpar M A 1985  {\it Nature} {\bf 316} 27



\bibitem{zuo:2004plb} Zuo W, Li Z H, Lu G C, Li J Q, Scheid W, Lombardo U, Schulze H J, Shen C W 2004 {\it Phys. Lett.} {\bf B595} 44


\bibitem{zhou:2004} Zhou X R, Schulze H J, Zhao E G, Pan F and Draayer J P 2004
{\it Phys. Rev.} {\bf C 70} 048802


\end{thebibliography}
\end{document}